\documentclass[12pt]{article}
\usepackage{epsfig}
\def\be{\begin{equation}}
\def\ee{\end{equation}}
\def\bea{\begin{eqnarray}}
\def\eea{\end{eqnarray}}
\usepackage{graphicx}

\catcode`\@=11
\def\lsim{\mathrel{\mathpalette\@versim<}}
\def\gsim{\mathrel{\mathpalette\@versim>}}
\def\@versim#1#2{\vcenter{\offinterlineskip
\ialign{$\m@th#1\hfil##\hfil$\crcr#2\crcr\sim\crcr } }}
\catcode`\@=12
\usepackage{axodraw}

\catcode`@=12
\begin{document}
\thispagestyle{empty}
\begin{flushright}
UCRHEP-T531\\
CETUP2013-007\\
November 2013\
\end{flushright}
\vspace{0.6in}
\begin{center}
{\LARGE \bf Unified Framework for Matter, Dark Matter,\\
and Radiative Neutrino Mass\\}
\vspace{1.2in}
{\bf Ernest Ma\\}
\vspace{0.2in}
{\sl Department of Physics and Astronomy, University of California,\\
Riverside, California 92521, USA\\}
\end{center}
\vspace{1.2in}
\begin{abstract}\
The well-studied radiative model of neutrino mass through $Z_2$ dark matter 
is shown to be naturally realizable in the context of $SU(6)$ grand 
unification.  A recent new proposal based on $U(1)_D$ dark matter is 
similarly accommodated in $SU(7)$.  Just as the proton is unstable at 
the scale of quark-lepton unification, dark matter is expected to be 
unstable at a similar scale.

\end{abstract}

\newpage
\baselineskip 24pt

The Universe consists of matter, dark matter, and dark energy.  Of the three, 
matter is understood in terms of the standard model (SM) of quarks and 
leptons with vector gauge bosons as force carriers and one fundamental 
physical scalar particle, i.e. the Higgs boson, which is presumably the 
126 GeV particle recently discovered~\cite{atlas12,cms12} at the Large 
Hadron Collider (LHC).  Whereas quark and charged-lepton masses are 
directly linked to the Higgs doublet $\Phi = (\phi^+,\phi^0)$, the origin 
of neutrino mass~\cite{m09} 
is not as clearcut.  It may involve physics beyond the SM and as such, may be 
linked to dark matter (as well as the Higgs boson) in a simple one-loop 
mechanism~\cite{m06} as shown below.  
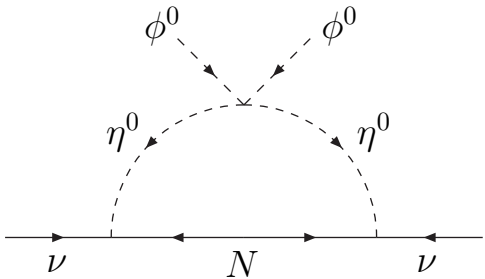
\begin{figure}[htb]
\begin{center}
\begin{picture}(360,120)(0,0)
\ArrowLine(90,10)(130,10)
\ArrowLine(180,10)(130,10)
\ArrowLine(180,10)(230,10)
\ArrowLine(270,10)(230,10)
\DashArrowLine(155,85)(180,60)3
\DashArrowLine(205,85)(180,60)3
\DashArrowArc(180,10)(50,90,180)3
\DashArrowArcn(180,10)(50,90,0)3

\Text(110,0)[]{\large $\nu$}
\Text(250,0)[]{\large $\nu$}
\Text(180,0)[]{\large $N$}
\Text(135,50)[]{\large $\eta^0$}
\Text(230,50)[]{\large $\eta^0$}
\Text(150,90)[]{\large $\phi^{0}$}
\Text(217,90)[]{\large $\phi^{0}$}

\end{picture}
\end{center}
\caption{One-loop generation of neutrino mass with $Z_2$ symmetry.}
\end{figure}
This well-studied model is very simple.  Three new singlet neutral Majorana 
fermions $N_{1,2,3}$ and one new scalar doublet $(\eta^+,\eta^0)$ are added to 
the SM with an exactly conserved discrete $Z_2$ symmetry, under which the new 
particles are odd and the SM particles are even.  It has been called 
'scotogenic' from the Greek 'scotos' meaning darkness.  Let $\eta^0 = 
(\eta_R + i \eta_I)/\sqrt{2}$, then the allowed $(\Phi^\dagger \eta)^2$ 
term splits the masses 
of $\eta_{R,I}$ so that the lighter, say $\eta_R$, is a good dark-matter 
candidate~\cite{m06}. Its detailed study began in Ref.~\cite{lnot07} and 
the most recent update is Ref.~\cite{kyr13}.

A new variation of Fig.~1 has recently been proposed~\cite{mpr13} as 
shown below.
\begin{figure}[htb]
\begin{center}
\begin{picture}(360,120)(0,0)
\ArrowLine(90,10)(130,10)
\ArrowLine(130,10)(180,10)
\ArrowLine(180,10)(230,10)
\ArrowLine(270,10)(230,10)
\DashArrowLine(155,85)(180,60)3
\DashArrowLine(205,85)(180,60)3
\DashArrowArc(180,10)(50,90,180)3
\DashArrowArcn(180,10)(50,90,0)3

\Text(110,0)[]{\large $\nu$}
\Text(250,0)[]{\large $\nu$}
\Text(155,0)[]{\large $N_{R}$}
\Text(205,0)[]{\large $N_{L}$}
\Text(135,50)[]{\large $\eta_1^0$}
\Text(230,50)[]{\large $\eta_2^0$}
\Text(150,90)[]{\large $\phi^{0}$}
\Text(217,90)[]{\large $\phi^{0}$}

\end{picture}
\end{center}
\caption{One-loop generation of neutrino mass with $U(1)_D$ symmetry.}
\end{figure}
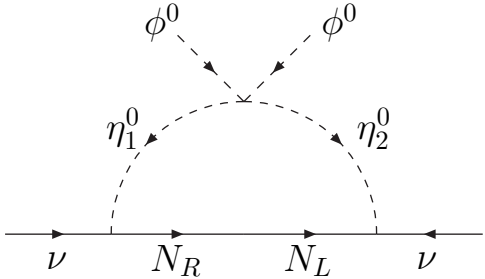
Instead of $Z_2$, it supports an $U(1)_D$ gauge symmetry which may be exact 
or broken, either into $Z_2$ or a conserved global U(1).  Here there are 
three new singlet neutral Dirac fermions $N_{1,2,3}$ and two new scalar 
doublets $(\eta^+_{1,2}, \eta^0_{1,2})$.  Whereas $N_{1,2,3}$ and $\eta_1$ 
transform as +1 under $U(1)_D$, $\eta_2$ transforms as $-1$.  The loop 
is thus completed without breaking $U(1)_D$.  Now the lightest $N$ is 
Dirac fermion dark matter, and the $U(1)_D$ gauge boson $\gamma_D$ and 
the corresponding Higgs boson $h_D$ are possible light force carriers 
of the self-interacting dark matter, with important astrophysical 
implications, such as the dark matter halo structure~\cite{tyz13} of 
dwarf galaxies and the possibility of Sommerfeld enhancement~\cite{lwz13} 
of dark matter annihilation to explain the observation of positron 
excess in recent space experiments~\cite{pamela09,ams13}.

What could be the origin of these new particles and new symmetries? 
It is proposed in this paper that they are remnants of a unified theory 
incorporating both matter and dark matter.  The notion that the observed 
quarks and leptons may be unified in a single theory, the first of which 
was $SU(5)$~\cite{gg74}, should be extended.  In the context of 
quark-lepton unification, the stability of the proton is not absolute. 
Its lifetime is very much longer than that of the Universe because the 
unification energy scale is very high, say $\sim 10^{16}$ GeV.  In the 
context of the unfication of matter and dark matter, it is thus expected 
that the $Z_2$ or $U(1)_D$ symmetry which maintains the stability of dark 
matter in the Universe today is again not absolute.  They will also be 
violated at a similar high energy scale.   To implement this simple idea, 
two prototype models based on $SU(6)$ and $SU(7)$ respectively will be 
discussed.  The former leads to the well-known original one-loop radiative 
(scotogenic) model of neutrino mass~\cite{m06} with $Z_2$ symmetry, as 
shown in Fig.~1, and the latter to the recent new variation~\cite{mpr13} 
with $U(1)_D$ gauge symmetry, as shown in Fig.~2.

The symmetry $SU(6)$ was originally considered~\cite{ikn77,y78} as an 
alternative to $SU(5)$ unification~\cite{gg74} with very different 
particle structure.  Here it is simply used as an extension of $SU(5)$ to 
include dark matter.  The well-known $SU(5)$ assignments of quarks and 
leptons are
\begin{equation}
\underline{5}^*_F = \pmatrix{d^c \cr d^c \cr d^c \cr e \cr \nu}, ~~~~~ 
\underline{10}_F = \pmatrix{0 & u^c & -u^c & -u & -d \cr 
-u^c & 0 & u^c & -u & -d \cr u^c & -u^c & 0 & -u & -d \cr 
u & u & u & 0 & -e^c \cr d & d & d & e^c & 0}.
\end{equation}
The standard-model Higgs doublet is contained in 
\begin{equation}
\underline{5}_S = \pmatrix{\xi^{-1/3} \cr \xi^{-1/3} \cr \xi^{-1/3} \cr 
\phi^+ \cr \phi^0},
\end{equation}
where the scalar color triplet $\xi^{-1/3}$ is assumed very heavy.  The 
resulting Yukawa couplings invariant under $SU(5)$ are
\begin{equation}
\underline{5}_F^* \times \underline{10}_F \times \underline{5}_S^*, ~~~~~ 
\underline{10}_F \times \underline{10}_F \times \underline{5}_S.
\end{equation}
As $\phi^0$ develops a nonzero vacuum expectation value, all quarks 
and leptons (except the neutrinos) obtain masses.

The $SU(6)$ extension is straightforward.  A neutral singlet fermion $N$ is 
added to $\underline{5}_F^*$ to form $\underline{6}_F^*$ under $SU(6)$. 
A heavy $\underline{5}_F$ is added to $\underline{10}_F$ to form 
$\underline{15}_F$ under $SU(6)$.  The analogous Yukawa couplings 
invariant under $SU(6)$ are
\begin{equation}
\underline{6}_F^* \times \underline{15}_F \times \underline{6}_S^*, ~~~~~ 
\underline{15}_F \times \underline{15}_F \times \underline{15}_S,
\end{equation}
where $\underline{6}_S^*$ contains $\underline{5}_S^*$ and $\underline{15}_S$ 
contains another $\underline{5}_S$.  Hence two different Higgs doublets are 
needed, one coupling to the $down$ sector, the other to the $up$ sector. 
Note that whereas $\underline{5}_F^* + \underline{10}_F$ is anomaly-free 
in $SU(5)$, the corresponding combination in $SU(6)$ is $\underline{6}_F^* 
+ \underline{6}_F^* + \underline{15}_F$.  The $\underline{5}_F^*$ 
contained in the extra $\underline{6}_F^*$ is heavy and pairs up with 
the $\underline{5}_F$  contained in the $\underline{15}_F$ already mentioned, 
through the $\underline{1}_S$ of $\underline{6}_S^*$.

The extension of $SU(6)$ to include dark matter has also been considered 
recently~\cite{b12} in a different context.  There the motivation is to 
understand the relative abundance of matter versus dark matter.  It is 
assumed that they are both generated with an asymmetry in the early 
Universe from a common source.  This scenario is very different from 
the one here which emphasizes the possible connection between neutrino 
mass and dark matter.

So far in the present case, $N$ is assumed to be the only new fermion at 
low energy, but it 
has no interaction with the SM particles.  Consider now 
the $SU(6)$ scalar multiplet $\underline{21}_S$.  It decomposes 
into $\underline{15}_S + \underline{5}_S + \underline{1}_S$ of $SU(5)$. 
The resulting Yukawa and quartic couplings invariant under $SU(6)$ are
\begin{equation}
\underline{6}_F^* \times \underline{6}_F^* \times \underline{21}_S, ~~~~~ 
\underline{15}_S^* \times \underline{15}_S^* \times \underline{21}_S \times 
\underline{21}_S. 
\end{equation}
These interactions generate three terms. (1) As the scalar component 
$\underline{1}_S$ of $\underline{21}_S$ acquires a vacuum expectation 
value, $N$ gets a Majorana mass. (2) There is now the interaction 
term $(\nu \eta^0 - e \eta^+)N$, where the scalar doublet $(\eta^+,\eta^0)$ 
comes from $\underline{5}_S$ of $\underline{21}_S$.  (3) There is also the 
quartic scalar coupling $(\Phi^\dagger \eta)^2$.  These three terms support 
a $Z_2$ symmetry under which $N$ and $\eta$ are odd and all others are 
even, thus realizing the existence of stable dark matter.  They are also 
exactly the terms required for the scotogenic 
neutrino mass~\cite{m06} of Fig.~1.   Its $SU(6)$ decomposition is 
shown in Fig.~3. 
\begin{figure}[htb]
\begin{center}
\begin{picture}(360,120)(0,0)
\ArrowLine(90,10)(130,10)
\ArrowLine(180,10)(130,10)
\ArrowLine(180,10)(230,10)
\ArrowLine(270,10)(230,10)
\DashArrowLine(155,85)(180,60)3
\DashArrowLine(205,85)(180,60)3
\DashArrowArc(180,10)(50,90,180)3
\DashArrowArcn(180,10)(50,90,0)3

\Text(110,0)[]{\large $6^*$}
\Text(250,0)[]{\large $6^*$}
\Text(150,0)[]{\large $6^*$}
\Text(210,0)[]{\large $6^*$}
\Text(180,0)[]{\large $21$}
\Text(180,10)[]{\large $\times$}
\Text(135,50)[]{\large $21$}
\Text(230,50)[]{\large $21$}
\Text(150,90)[]{\large $15$}
\Text(217,90)[]{\large $15$}

\end{picture}
\end{center}
\caption{$SU(6)$ decomposition of Fig.~1.}
\end{figure}
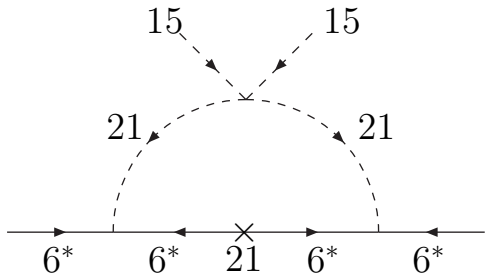

However, the $Z_2$ symmetry here is not absolute.  Just as the heavy 
color triplet gauge bosons contained in the adjoint $\underline{24}_V$ of 
$SU(5)$ mediate proton decay, the extra heavy color triplet gauge bosons 
contained in the adjoint $\underline{35}_V$ of $SU(6)$ will connect $N$ to 
the quarks, resulting in $N$ decaying into $p \pi^-$ and $n \pi^0$.  
Another possibility is the mixing of the heavy scalar color triplet 
$\zeta^{-1/3}$ in $\underline{21}_S$ with its counterpart $\xi^{-1/3}$ in 
$\underline{15}_S$ through the adjoint $\underline{35}_S$ of $SU(6)$. 
The companion interactions $d^c \zeta N$ together with $u d \xi$ from 
$\underline{15}_F \times \underline{15}_F \times \underline{15}_S$ imply also 
the decay $N \to p \pi^-$ or $n \pi^0$, just as the proton decays through 
the heavy scalar color triplet $\xi$ itself.  A unified 
framework for understanding matter and dark matter is thus established 
together with the origin of neutrino mass.

In the breaking of $SU(6)$ through the adjoint $\underline{35}_S$, let the 
$6 \times 6$ matrix representing $\underline{35}_S$ develop diagonal 
vacuum expectation values proportional to $(1,1,1,0,0,-3)$, then the 
residual symmetry is $SU(3)_C \times SU(2)_L \times U(1)_Y \times U(1)_N$. 
As noted earlier, $\underline{21}_S$ breaks $U(1)_N$ at a lower energy 
scale for $N$ to acquire a Majorana mass.  The above choice of 
$\underline{35}_S$ breaking mixes $\zeta$ with $\xi$, but not $\eta$ with 
$\Phi$.  It is only possible here because of $SU(6)$.  If the gauge group 
is $SU(5)$, then the breaking to $SU(3)_C \times SU(2)_L \times U(1)_Y$ 
must be along the $(1,1,1,-3/2,-3/2)$ direction and that will mix both 
sectors.

Of course, this simple $SU(6)$ model is still missing many details, such as 
the complete fermion and scalar content necessary to obtain the requisite 
low-energy particle spectrum with the effective $Z_2$ symmetry and how 
gauge couplings may be unified, with or without supersymmetry, among many 
other possible issues.  On the other 
hand, the important point has been made that the stability of dark matter 
may be linked to the stability of ordinary matter in much the same way.
Such is the intended message of this paper.

Consider next the symmetry $SU(7)$~\cite{cyc80,uy81}.  In the decomposition 
$SU(5) \times SU(2)_N$, the $\underline{7}_F^*$ of $SU(7)$ contains an extra 
$SU(2)_N$ doublet $(N_1,N_2)$.  The corresponding Yukawa couplings to 
those of Eqs.~(3) and (4) are
\begin{equation}
\underline{7}_F^* \times \underline{21}_F \times \underline{7}_S^*, ~~~~~ 
\underline{21}_F \times \underline{21}_F \times \underline{35}_S. 
\end{equation}
The $\underline{28}_S$ of $SU(7)$ now contains a bidoublet, i.e. 
\begin{equation}
\pmatrix{\eta_1^+ & \eta_2^+ \cr \eta_1^0 & \eta_2^0},
\end{equation}
where $SU(2)_L$ applies vertically and $SU(2)_N$ applies horizontally.
The corresponding couplings to Eq.~(5) are
\begin{equation}
\underline{7}_F^* \times \underline{7}_F^* \times \underline{28}_S, ~~~~~ 
\underline{21}_S^* \times \underline{21}_S^* \times \underline{28}_S 
\times \underline{28}_S. 
\end{equation}
As a result, Fig.~2 is obtained, with $U(1)_D$ given by the diagonal subgroup 
of $SU(2)_N$.  Its $SU(7)$ decomposition is shown in Fig.~4. 
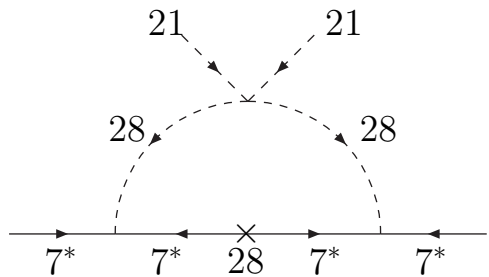
\begin{figure}[htb]
\begin{center}
\begin{picture}(360,120)(0,0)
\ArrowLine(90,10)(130,10)
\ArrowLine(180,10)(130,10)
\ArrowLine(180,10)(230,10)
\ArrowLine(270,10)(230,10)
\DashArrowLine(155,85)(180,60)3
\DashArrowLine(205,85)(180,60)3
\DashArrowArc(180,10)(50,90,180)3
\DashArrowArcn(180,10)(50,90,0)3

\Text(110,0)[]{\large $7^*$}
\Text(250,0)[]{\large $7^*$}
\Text(150,0)[]{\large $7^*$}
\Text(210,0)[]{\large $7^*$}
\Text(180,0)[]{\large $28$}
\Text(180,10)[]{\large $\times$}
\Text(135,50)[]{\large $28$}
\Text(230,50)[]{\large $28$}
\Text(150,90)[]{\large $21$}
\Text(217,90)[]{\large $21$}

\end{picture}
\end{center}
\end{figure}

In conclusion, it has been shown that a unified framework of matter and 
dark matter is possible and perhaps even desirable for the complete 
understanding of our Universe.  Two prototype models based on $SU(6)$ 
and $SU(7)$ have been discussed.  The former applies naturally to the 
original scotogenic model of neutrino mass with $Z_2$ symmetry.  The 
latter fits in nicely with the recent proposal of a possible exact 
or broken $U(1)_D$ symmetry in place of $Z_2$.  This new dynamical $U(1)_D$ for 
dark matter has many impications for astrophysical observations.

\noindent \underline{Acknowledgment}~:~I thank CETUP$^*$ (Lead, South Dakota) 
for hospitality 
and providing a stimulating environment where the main idea of this paper 
was conceived. This work is supported in part 
by the U.~S.~Department of Energy under Grant No.~DE-SC0008541.

\bibliographystyle{unsrt}

\end{document}